# MOMENTS IN THE PRODUCTION OF SPACE: DEVELOPING A GENERIC ADOLESCENT GIRLS AND YOUNG WOMEN HEALTH INFORMATION SYSTEM IN ZIMBABWE

Rangarirai Matavire, Itinordic AS, matavirer@gmail.com

Jørn Braa, University of Oslo, jbraa@ifi.uio.no

Shorai Huwa, Zimbabwe National AIDS Council, shuwa@nac.org.zw

Lameck Munangaidzwa, Zimbabwe National AIDS Council, lmunangaidzwa@nac.org.zw

Zeferino Saugene, Eduardo Mondlane University, zsaugene@gmail.com

Isaac Taramusi, Zimbabwe National AIDS Council, itaramusi@nac.org.zw

Bob Jolliffe, University of Oslo, bobjolliffe@gmail.com

**Abstract:** With global targets to end AIDS by 2030 and to eliminate new HIV infections, Adolescent Girls and Young Women (AGYW) are seen to be particularly vulnerable, especially in Sub Saharan Africa. Numerous nations have therefore rolled out interventions to provide services to remove the determinants of vulnerability, such as limited education, early marriage, poverty, domestic violence, and exposure by male partners. Within this context, subpopulations such as sex workers increase the vulnerability amongst AGYW and are also supported through prevention programming. This study follows a project to develop a generic health information systems solution to provide a means to monitor and evaluate the successes of the AGYW initiative in reducing new infections. It borrows theoretical ideas from Henri Lefebvre's theory of moments to describe the process in which the space for the development of the solution is produced.

## 1. INTRODUCTION

With AIDS-related illness continuing to be among the leading causes of death and the social impact of new and existing HIV infections evident, particularly amongst women in under-privileged societies, recent advances in medicine have spawned an emergent vision to end AIDS by 2030. One such intervention aims to reduce and eliminate new infections among Adolescent Girls and Young Women (AGYW). According to UNAIDS (2019), "*three in four new HIV infections among 10–19-year-olds are among girls*" with a markedly high incidence amongst Key Population (KP) groups such as sex workers, and prisoners. The government of Zimbabwe, through the National AIDS Council (NAC) is part of a 13 nation Sub-Saharan Africa initiative aimed at reducing HIV incidence among AGYW. AGYW are classified as females between the ages of 15 and 24 years of age, which includes their male partners (UNAIDS, 2016). NAC was established by an act of parliament in 1999 and is responsible for coordinating the national multi-sectoral response to HIV and AIDS in Zimbabwe.

NAC monitors a number of HIV prevention programs, many of which the organisation has a coordinating role at the district level. Below the districts, several implementing partners are coordinated by NAC to provide services at the community level. The implementing partners are a group of organisations, both private and public, who implement NAC's programs across the 60+

districts in Zimbabwe. NAC itself implements prevention programs directly in some of these districts. In addition, NAC also operates within the broader context of the Ministry of Health and Child Care (MoHCC) which runs 1600+ health facilities nationwide and provides HIV services such as testing, counselling, treatment, and other community interventions in the respective catchment areas. Within this context, some of the implementation partners are directly involved in initiatives specifically targeting AGYW, such as the Determined, Resilient, Empowered, AIDS-free, Mentored and Safe (DREAMS) program.

### 1.1. The AGYW Initiative

#### 1.1.1. Background

AGYW in Zimbabwe are 14 times more likely to get a new HIV infection than their male counterparts (Saul, Bachman, Allen, Toiv, Cooney & Beamon, 2018). As a component of a broader global goal to end AIDS by 2030, AGYW interventions are constituted of a set of partnerships aimed at reducing the incidence of new HIV infections amongst females who are 15-24 years old through a multi-pronged and multi-stakeholder approach. The partnerships target the determinants of new HIV infections such as poverty, inequality, biomedical and structural issues. This is done by first identifying girls in a community within this age group, assessing individual risk and subsequently enrolling vulnerable girls for age-appropriate interventions. As highlighted earlier, NAC's main role is to coordinate the HIV and AIDS activities of the multiple stakeholders operating in Zimbabwe. NAC itself is implementing a number of prevention programs in some districts. Among these are initiatives aimed at subpopulations which have generally higher incidence of HIV. For example, female sex workers in Sub Saharan Africa have up to 21 times greater incidence of the infection than the general population (UNAIDS, 2019). Female sex workers, as with other Key Populations (KPs), also constitute a subpopulation that can be found among AGYW.

A cornerstone of the intervention is a functioning Monitoring and Evaluation (M&E) system for clinical decision support at the service point and for calculation of indicators (measures) at different points of the hierarchy for decision making. At the onset of this work, there was no integrated M&E system to provide reports on globally reported indicators for AGYW in NAC, and this was the basis for the project described in this paper. However, as mentioned earlier, other organisations had implemented the DREAMS initiative in some districts under an International non-governmental organisation called INGO. This study therefore constitutes the moments in a process to scale up the DREAMS work nationwide, under NAC with the added aim to develop a generic AGYW system. With this background, the initiative under the watch and implementation of NAC has in some discussions been referred to as the Modified DREAMS initiative. A background on DREAMS, which constituted the installed base in Zimbabwe, is given below. The installed base is the pre-existing configuration of people, technologies, process, and systems (Hornnes, Jansen & Langeland, 2010).

#### 1.1.2. Determined, Resilient, Empowered, AIDS-free, Mentored and Safe

Given the complex interplay of factors involved, intervention by a single player cannot be effective in reducing the incidence of new HIV infections among AGYW (Saul et al., 2018). To address this issue, DREAMS consists of a package of interventions targeted at the individual girl or woman, her family, the community, and her male partner(s). The package consists of biomedical, social and economic interventions which include services in the areas of educational subsidies, HIV testing and counseling, parenting/caregiver programs, domestic violence reduction and male partner circumcision among others. Zimbabwe is among the 10 African countries which began the DREAMS initiative in 2015 given that new HIV infections among AGYW from these nations constitute about half of those occurring globally. A key aspect of the DREAMS initiative in Zimbabwe was its use of the District Health Information Software (DHIS2) for its case-based surveillance. This software has also been in use as the national HMIS for routine data since 2012 and for case-based surveillance in a number of programs such as for the national malaria program since 2014 (Matavire, 2016).

### 1.2. Research questions and goals

This study attempts to describe and explain the process of evolution of the health information system by highlighting the global and local collaborative efforts and their results through different phases. The aim of the study is to theoretically step back from the actual health information system as a product, and to understand the social relationships through which it emerges. This is in line with Lefebvre and Nicholson-Smith (1991, p. 113) who contend that, "*it is never easy to get back from the object (product or work) to the activity that produced and/or created it. It is the only way, however, to illuminate the object's nature, or, if you will, and reconstitute the process of its genesis and the development of its meaning*". It is in this regard that our primary research question is, 'How can we understand the activities implicated in the development of the generic AGYW system in Zimbabwe?'. The theoretical goal is to reveal the underlying struggles in the process of systems development and implementation with the intention of creating possibilities for users to appropriate the technology. This goal contrasts with merely imposing systems on behalf of dominating actors and constellations to the detriment of users, developers and implementers (Read, de Laat-Lukkassen, & Jonauskis, 2013). It is in this context that Henri Lefebvre's theory on the '*production of space*' is found to be a fertile ground for the development of a theoretical and practical understanding of the phenomenon under study, as will be demonstrated in the proceeding sections.

## 2. THEORETICAL BACKGROUND

There is an understanding that most health information systems (HIS) initiatives fail in some way (Heeks, 2006). This extends beyond the realm of HIS, even beyond the IT world, to fields such as sociology, economics, politics and philosophy where questions of what the world is constituted of, that is ontology, and what constitutes knowledge, that is epistemology, are pursued. Notwithstanding, we posit that many studies in the IS discipline overlook the interplay between mental, social and physical spaces as a basis to understanding success and failure of interventions (Matavire, 2016). Instead, there has been a proliferation of studies which use theories such as Structuration Theory (Orlikowski, 1991) and Actor Network Theory (Braa, Monteiro & Sahay, 2004), among others, where for example the relationship between agency and structure, and the subjectivity of the agent are scarcely problematised, respectively. In this sense, information systems researchers have not adequately illuminated the way knowledge serves as an instrument in the hands of power. It is within this context that Lefebvre's theory on the 'production of space' (Lefebvre & Nicholson-Smith, 1991) is particularly exciting in trying to describe the phenomenon of HIS production.

### 2.1. Production of space

In general, what Lefebvre states is that the concept of space is increasingly important in understanding everyday life. He argues that while researchers utilise the word space in their descriptions, they take it for granted by often privileging one space, often the mental, over all others such as physical and social, and vice versa. In this way, much of the domination of one space over the others is not only taken for granted, but rather is deliberately concealed. A core concept in understanding the production of space is 'abstract space'. This is a space that is produced by capitalism and neo-capitalism and includes "*the 'world of commodities', its logic and its worldwide strategies, as well as the power of money and that of the political state. This [abstract] space is founded on the vast network of banks, business centres and major productive entities, as also on motorways, airports and information lattices*" (Lefebvre & Nicholson-Smith, 1991, p. 53). Consequently, Lefebvre developed a theory of space to explain and describe the genesis of spaces, such as this abstract space and other myriads of spaces in everyday life.

To Lefebvre, space is a product of the pervading mode of production. In the contemporary world, capitalism therefore produces its own space. In this regard, other alternative modes of production such as socialism consequently have not produced their space, and hence have not been successful projects. In contrast, differentiated spaces have emerged to counter the abstract homogenised space of capitalism, yet still within it. In this context, some societies produce their own space, to counter the homogenising effects of the abstract space of capital, and to signify its impermanence, through class struggle. In the context of technology development, it is therefore possible to look beyond the artifact, and understand the relationships through which its space is produced. To understand this, Lefebvre develops a theory for the moments of production as elaborated on below.

### 2.1.1. Moments in the production of space

Every society produces its own space, or else it remains in the level of ideas with no concrete existence (Lefebvre & Nicholson-Smith, 1991). Space is produced through the moments of i) *spatial practice (perceived)*, which denotes the reproduction of the social relations of production, ii) *representations of space (conceived)* which are "*in thrall to both knowledge and power*" (Lefebvre & Nicholson-Smith, 1991, p. 50), and iii) *representational spaces (lived)* which are limited to works and other symbolic aspects of life. Lefebvre makes a distinction between product and work, where a work is related to art and product is related to reproduction. Conceived space is the space of planners such as the producers of standards and generic software tools; Perceived space is that which subsumes the social relationships through which activities proceed; Lived space is the actual social practices, everyday use and life of a space which somewhat escapes grasp of planners as it is constituted of meaning, beliefs and superstition. Rather than seeing these in isolation, Lefebvre sees the three as constituting the 'moments' in the production of space (Lefebvre & Nicholson-Smith, 1991).

## 3. METHODOLOGY

In essence, this study is part of an ongoing series of work which started with the implementation of the DHIS2 system in the health ministry of Zimbabwe in 2012. Underlying this work is an action-oriented approach inspired by the Scandinavian tradition of Action Research (Baskerville, 1999). The study is also influenced by Grounded Theory methodology (GTM) (Glaser & Strauss, 1967), albeit not in the sense of a rigorous application of its principles such as coding techniques, or the role of theory, but in the sense that the study aims to develop theory which can be applicable across diverse terrains where similar technologies are implemented. In this context, Eisenhardt's (1989) ideas on developing grounded theory from case studies are adopted in the study. The combination of principles from diverse research methods, so called mixed methods, is not novel to this study but is in fact common in information systems research (Matavire & Brown, 2013; Baskerville & Pries-Heje, 1999).

There are two meanings to the term "grounded theory", the first being the result of a study and the second being the methodology. In the first, a 'grounded theory' is the conceptual outcome of a study, that is the theory that is developed through a known procedure or technique. In the second, we refer to the methodology as developed originally by Glaser and Strauss (1967), with its subsequent erosion and/or elucidation through further works such as Strauss and Corbin's (1997) Straussian GTM, Charmaz's (2006) constructivist GTM, Baskerville and Pries-Heje's (1999) or Eisenhardt's (1989) Mixed GTM, and the use of analytical techniques from GTM such as open coding without adopting the tenets of any one flavour such as is common in information systems research (Matavire & Brown, 2013). In this context, this study aims to reflexively (Alvesson & Sköldberg, 2000) develop a 'grounded theory' to explain the researched phenomenon by using a case study oriented approach as envisaged by Eisenhardt (1989), through participation and action (Baskerville & Pries-Heje, 1999). All the authors of this paper have been actively involved in the development and implementation of the system.

The case is developed through an analysis of various source documents and communications which have characterised the intervention from system conceptualisation to production. These include documents from global partners, data entry forms, discussions among stakeholders, meetings with developers, user feedback, contracts, project reports, and various ceremonies and occasions marking transitions in the process of development. As mentioned previously, the researchers are themselves part of the initiative, consequently reflections on the journey, with necessary clarifications sought from the stakeholders, with their divergent and convergent interests, have contributed to the emergence of the case. A hermeneutic approach (Klein & Myers, 1999) to understanding Henri Lefebvre's theory of moments has been important in constructing the case. There has consequently been an iterative process involved in the construction of the case in relation to the chosen theory. The authors collectively have a rich understanding of the empirical domain, consequently a theoretical framework has been necessary for gleaning the case from the source material and the experiences of the researchers. The suitability of Henri Lefebvre's theory on the 'production of space' to explain the phenomenon in health information systems is part of an ongoing project for some of the authors, yet the focus on 'moments' has been novel to this study and marks a key point in the development of the framework.

## 4. THE CASE OF THE DHIS2 AGYW/KP INITIATIVE

An international NGO, which we can refer to as INGO, was tasked with developing a Monitoring and Evaluation (M&E) system for the DREAMS program running in a few districts in Zimbabwe prior to 2018. In this context, the DHIS2 Tracker module was customised by INGO for their 'DREAMS' project which targeted various AGYW HIV/AIDS services and providers in the community. DHIS2 is a customisable web-based software that enables the collection of both aggregate and case-based health data across regions for storage and analysis in a centralised repository at local, administrative, and central levels as illustrated in Figure 1., below. The system can also capture data at a community level offline on mobile devices. It is widely used in the Global South, with over 80 nationwide implementations. DHIS2 Tracker is the case-based module within the system which enables the collection of individual level data that might include identifiable information like names and surnames, and event-based data such as encounters at the facility, in a manner similar to a typical Electronic Medical Record (EMR) system. It was this module that was implemented by INGO to track and monitor AGYW interventions across providers.

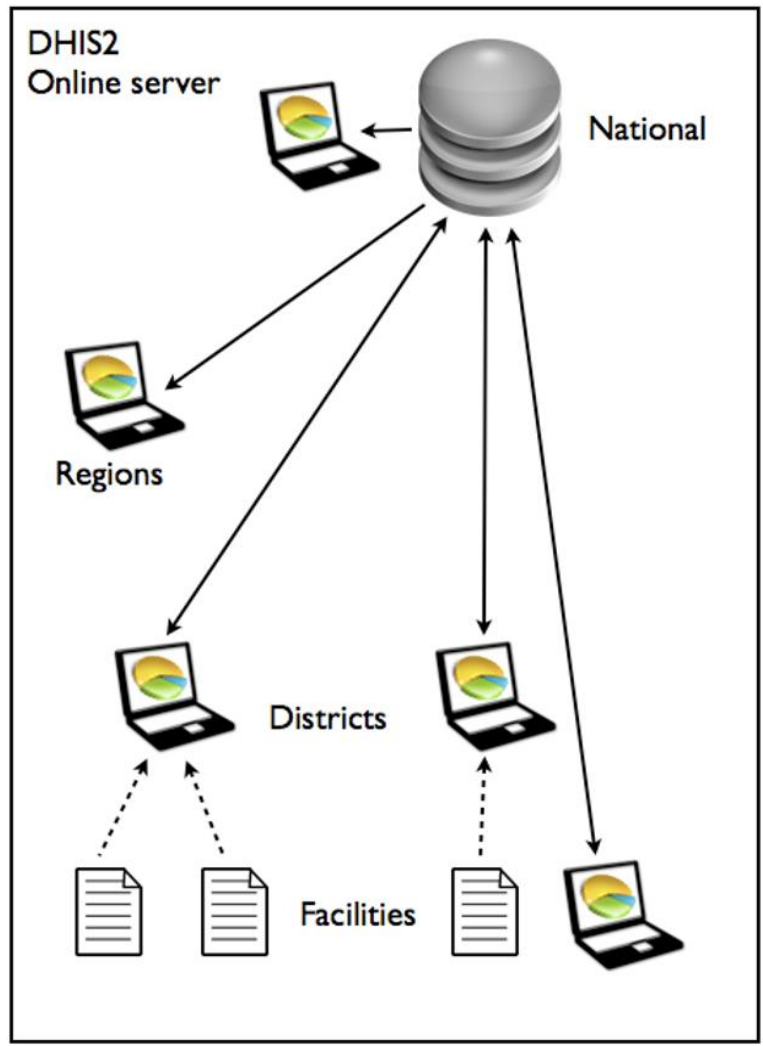

**Figure 1.Information flows in DHIS2**

### 4.1. Unique Identification of AGYW

Within many developing nations, there is a *wicked problem* of unique identification of beneficiaries of services given that there is often no fully functional nationwide identity system in place. A key innovation in INGO's implementation was the issue of a reliable client identifier. AGYW is implemented across services and providers, at the community level, where unique identification of clients can be difficult, also given the offline collection of information which often occurs on paper forms. In this context, a client who is registered for a particular service, let us say an 'educational subsidy' under provider X, needed to be uniquely identified if they accessed another service such as 'HIV counselling and testing' at a government facility. This is part of a layering of services within a package so as to permit the generation of indicators such as '*number of girls who received an educational subsidy who also had a negative HIV result*' within a period. To address this pertinent issue, INGO created a demographic identification system, which can be generated on paper and offline using some predefined rules and applied across different providers and for different services. The system works by taking parts of a client name, surname, their contact, date of birth, gender and gender identity to construct a 11-character identifier for the beneficiary. While in some cases the same identifier can be produced for more than one beneficiary, it has been largely successful in unique identification, with some modifications in progress to decrease the possibility of clashes.

## 4.2. From DREAMS to a Generic AGYW product

In 2018, recognising the success of INGO in implementing the set of DREAMS packages in some districts for AGYW, NAC engaged its international partners and donors in order to scale up the initiative nationwide. At the global level, partners were also in discussion on how to scale up the initiatives into other countries. In this context, the Health Information Systems Program (HISP) in the University of Oslo was contracted to scale up the initiative in Zimbabwe with NAC and its partners. HISP was also tasked with developing generic features in the core of its software to better support the AGYW use case. In addition, HISP was tasked with also developing a set of information products, that is generic customisations, on its platform to help other countries to adapt and implement in their diverse contexts based on the Zimbabwe use case.

### 4.2.1. Collaboration agreements amongst partners

Initially, HISP developed a comprehensive proposal on the development and implementation of the system, with a focus on a generic AGYW solution, to be implemented in 30 of Zimbabwe's 60+ districts. INGO had previously existing agreements with HISP to support its activities in different use contexts on which they were actively collaborating. Consequently, there was a strategy with the HISP team to learn from, build upon, support, and scale up existing INGO work on the AGYW initiative in Zimbabwe in a collaborative manner. Numerous calls were scheduled between the stakeholders, at both local and global levels, to initiate the collaboration between INGO and HISP as implementers and developers of the Zimbabwe AGYW system. Also, the aim of the collaboration was to develop generic AGYW features in the core DHIS2 software which INGO would benefit from in scaling into other regions in which DREAMS was implemented.

At a global level, in July 2018, 4 international HISP consultants, including one from INGO who had designed the DREAMS DHIS2 implementation in Zimbabwe planned a mission to meet stakeholders and to conduct an analysis of the existing system and produce a plan of implementation. 3 of the consultants are co-authors of this paper. The mission included visiting the implementing districts in Zimbabwe to learn about the AGYW intervention. This involved meeting with community leaders, observing implementation of domestic violence support programs, visiting schools to understand AGYW educational support, meeting with Ministry of Health and Child Care (MoHCC) officials, and finalising the plan for presentation to responsible departmental heads. Discussions between HISP and INGO were also scheduled to agree on the roles of the different actors in the emergent implementation. Also given the existing collaboration between HISP and MoHCC which had begun in 2012 on the nationwide DHIS2 system, meetings and discussions were also held between the two organisations. The mission visit was concluded, and a preliminary plan and schedule had been developed, and roles agreed.

Globally INGO expressed an interest and commitment to be collaborating as partners with HISP as part of the existing arrangements. A key meeting between the local office of INGO, NAC, and HISP was held in March 2019. Before the meeting, the funding partner had shared the comprehensive proposal, plan and roadmap on the implementation of the Generic AGYW system which had been developed by HISP after the earlier field visit for discussion. Issues such as the hosting location of the existing DREAMS servers, the hosting arrangement of the new Generic AGYW implementation, the roles of the different organisations, such as the local INGO team in relation to the HISP team, timing and alignment of activities were discussed among the stakeholders. INGO in particular needed to consult with its international office as it had its own proposal, funding, scope and timing in the scaling up of DREAMS in the NAC intervention.

HISP updated the Global INGO office on the progress made in April 2019 where the local office of INGO also agreed to subsequently share documents on its own roadmap for expanding the DREAMS initiative within NAC. A multiplicity of interests needed to be aligned, both amongst local organisations and global collaborators in the different areas and periods of intervention. In addition, INGO had a pre-existing collaboration with NAC. The INGO roadmap was used as the basis of collaboration, with HISP working to integrate its work in the proposed timeline of activities.

The work of integrating the roadmap continued through May 2019. The component of the work to be conducted by HISP was shared with the funders of the project. A set of deliverables were agreed in the conduct of the work. These included tasks around integration, mentoring and training, system reviews, and the development of generic AGYW tools. Agreements on the objectives and roadmap led to the signing of the contracts in December 2019, paving the way for HISP's intervention to begin.

### 4.2.2. Re-aligning roles and scope

In February 2020, HISP organised its 1st mission on the project, with an objective to initialise the activities of the work on the ground. One of the consultants, also co-author on this paper, attended the inception meeting on behalf of HISP during this period. The meeting was coordinated between NAC, its local implementation partner, a United Nations agency (UNAgency), and HISP. Several points were discussed and agreed upon with the leadership of both UNAgency and NAC. The key point was that the customisation and set up of the system should proceed from analysis of the legacy system in NAC, not necessarily on the DREAMS configuration, and implemented in agreement with NAC M&E requirements onto the DHIS2 platform. This gave HISP a more prominent and demanding local role on the project than had been originally anticipated. There were now uncertainties on the scope of collaboration between INGO and HISP at the local level. Access to the pre-existing configuration remained important but limited through other moments in the evolution of the solution.

Focus shifted to pre-existing arrangements at the global level between INGO and HISP. A key deliverable for HISP to the funder was a generic AGYW solution, and INGO had already been involved with the global HISP to support specific features of their DREAMS work in Zimbabwe, among other countries. Within HISP, various actors were discussing with INGO on features that were important for analysing the layered data in DREAMS. The HISP team working on the NAC project joined into these conversations to facilitate development of generic AGYW. These were constituted of weekly meetings all the way into July of 2020. Meanwhile, the analysis and customisation of the NAC system was ongoing. A team had been created after the meeting earlier in February and had started collaborating on the customisation. Within this context, the Covid-19 pandemic had already begun, with all planned travel cancelled.

### 4.2.3. Customisation and piloting

A team consisting of M&E officers and IT staff from NAC and DHIS2 specialists from HISP was constituted following the stakeholders meeting in February 2020. NAC had been in the process of re-aligning their data entry formats and had ongoing planned activities in that vein. As soon as the redesigned data collection forms had been available, they were shared online with the HISP team for analysis and implementation. The initial version of the forms was shared in March, and the next in May 2020. The required customisation consisted of two main aspects, the first being routine aggregate disease surveillance, and the other being case-based surveillance. Within NAC, aggregated information is collected on a quarterly basis into a consolidated form. This includes aggregate data from diverse program areas working at community level and reporting to the districts. The customisation of these forms was conducted from March 2020, through to early 2021. The aggregated forms were also successfully piloted and by the time of this report, approaching rollout.

Meanwhile, as the aggregate surveillance system was being customised, work was also being undertaken to implement the data collection forms for 9 program areas which included support groups, key populations, gender and AGYW. Weekly calls were held with HISP, who were customising the system, and NAC, who were testing and providing feedback. With all programs created and workflows developed, the HISP team started to work on the generic AGYW solution with the internal DHIS2 developers. The concept was developed, based on experiences with the NAC customisation, and as part of it, into a generic AGYW product. This was also shared with NAC and the funder for comments. At the point of writing this article, customisation of the AGYW system, based on initial feedback and the availability of additional information, including the integration of the generic concept, was underway so the pilot could be initiated.

## 5. DISCUSSION

In our context, the generic AGYW solution was to be developed as an artifact for the Zimbabwe case that could be reproduced in another location. The space of representation (lived space) in the context of this project is constituted of the locations of the diverse stakeholders, both global and local, who through a process of alignment occurring in the abstract space, come together (physically and online) to define and agree on the scope and roadmap of the intervention, that is representations of space (conceived space). This is the space of the planners, and it is revealed in the initial step within the intervention, where alignment of goals, plans, is sought. In this first moment, a space for collaboration, itself pre-existing as abstract space, forms plans to define the scope of the intervention. This is a moment of planning, when actors in abstract space come together, and discuss the goals of the initiative, and the steps to realise it, and the people who will execute this plan. However, it must be mentioned that INGO, as a unilateral NGO, had underlying conflicts with UNAgency, and HISP which are generally multilateral. As illustrated, this space of representation is itself evolving, as the goals change, so is its constitution. In the context of this study, the moment of the space of representation has been largely dominant, with underlying conflict.

The moment of the spatial practice can be seen as at least constituted within the testing and feedback phase, where users of the system participate in the development process through testing and feedback. In this phase, users take the system into their locations and attempt to navigate across the system, learning about its operation through a mentorship approach. In learning the boundary of the system, applying known pathways to check on the fit to the context, there is an emergence of the spatial practice into clearer focus, that is the who's who of the initiative. As can be noted, the feedback between HISP and NAC to iteratively improve the technology constitutes a dialectic between the moment of the space of representation and the spatial practice. Lefebvre and Nicholson-Smith (1991) have in their work demonstrated the essential role of the dialectic between the different moments in the production of space. This also fosters a sense of ownership of the emergent spatial practice for the users.

In the case of the representational space, in lived space, the space of the users, a space of ethnographers, where meanings take stage, there is a sense that this was displaced and ineffectual. While users have been actively involved in the testing of the system, their day-to-day lived realities with the system were not adequately explored, also given some difficulties of travel during the study. The location of the actors in Zimbabwe and Norway, builds upon the pre-existing abstract space where the North develops the system and the South consumes it. However, with the case-based system emerged a need for mobile devices. Mobile devices can constitute another lived spatial moment of the users of the technology. Issues with electricity and connectivity as well require that the moment of lived space be adequately analysed in future.

## 6. CONCLUSION

We have applied Lefebvrerian concepts to analyse key moments in the production of a space for the development of a generic AGYW solution. In this instance, the moments of Lefebvre's spatial triad were applied in describing the process through which a generic health information system is produced. The theory of the production of space has enabled us to analyse the process in which a novel intervention is produced, and the reproductive moments which are essential for its emergence. The usefulness of Lefebvre in analysing such phenomena can be seen in the way in which his theory of moments reveals how mental ideas are brought into physical space to intervene and shape it, and indeed produce it. Using this conceptualisation we can question the success of the intervention by looking at what it is that is lacking to enable that a differential space is produced, not dominated by abstract space. Further work will aim to follow this intervention, to the point of it being moved from site to site, and how this impacts on the produced space. In addition, it will be important to delve deeper into the concept of production of space to find out what it can offer information systems research.

# REFERENCES AND CITATIONS

Alvesson, M., & Sköldberg, K. (2000). Reflexive methodology: New vistas for qualitative research. Sage.

Baskerville, R. L. (1999). Investigating information systems with action research. Communications of the Association for Information Systems, 2(1), 19.

Baskerville, R., & Pries-Heje, J. (1999). Grounded action research: a method for understanding IT in practice. Accounting, Management and Information Technologies, 9(1), 1-23.

Braa, J., Monteiro, E., & Sahay, S. (2004). Networks of action: sustainable health information systems across developing countries. MIS quarterly, 337-362.

Charmaz, K. (2006). Constructing grounded theory: A practical guide through qualitative analysis. Sage.

Eisenhardt, K. M. (1989). Building theories from case study research. Academy of management review, 14(4), 532-550.

Glaser, B. G., & Strauss, A. L. (1967). Discovery of grounded theory: Strategies for qualitative research. Routledge.

Heeks, R. (2006). Health information systems: Failure, success and improvisation. International journal of medical informatics, 75(2), 125-137.

Hirschheim, R., & Klein, H. K. (1989). Four paradigms of information systems development. Communications of the ACM, 32(10), 1199-1216.

Hornnes, E., Jansen, A., & Langeland, Ø. (2010, August). How to develop an open and flexible information infrastructure for the public sector?. In International Conference on Electronic Government (pp. 301-314). Springer, Berlin, Heidelberg.

Klein, H. K., & Myers, M. D. (1999). A set of principles for conducting and evaluating interpretive field studies in information systems. MIS quarterly, 67-93.

Lefebvre, H., & Nicholson-Smith, D. (1991). The production of space (Vol. 142). Blackwell: Oxford.

Matavire, R. (2016). Health Information Systems Development: Producing a New Agora in Zimbabwe. Information Technologies & International Development, 12(1), pp-35.

Matavire, R., & Brown, I. (2013). Profiling grounded theory approaches in information systems research. European Journal of Information Systems, 22(1), 119-129.

Orlikowski, W. J. (1992). The duality of technology: Rethinking the concept of technology in organizations. Organization science, 3(3), 398-427.

Read, S., de Laat-Lukkassen, M., & Jonauskis, T. (2013). Revisiting "complexification," technology, and urban form in Lefebvre. Space and Culture, 16(3), 381-396.

Saul J., Bachman G., Allen S., Toiv N. F., Cooney C. & Beamon T. (2018). The DREAMS core package of interventions: A comprehensive approach to preventing HIV among adolescent girls and young women. PLoS ONE 13(12): e0208167. https://doi.org/10.1371/journal.pone.0208167

Strauss, A., & Corbin, J. M. (1997). Grounded theory in practice. Sage.

UNAIDS (2019). At a glance. Retrieved from https://www.unaids.org/en/resources/infographics/women_girls_hiv_sub_saharan_africa

UNAIDS (2016). Guidance: HIV prevention among adolescent girls and young women Retrieved from https://www.unaids.org/sites/default/files/media_asset/UNAIDS_HIV_prevention_among_adolescent_girls_and_young_women.pdf